\documentclass[]{interact}
\usepackage{epstopdf}
\usepackage{color}
\usepackage[numbers,sort&compress]{natbib}
\usepackage{lipsum}
\usepackage{adjustbox}
\usepackage{amssymb}
\usepackage{amsmath} 
\usepackage{anyfontsize}
\usepackage{bbm}
\usepackage{enumerate}
\usepackage{bm}
\usepackage{color} 
\usepackage[dvipsnames]{xcolor}
\usepackage{enumerate}
\usepackage[font=scriptsize]{caption} 
\usepackage{graphicx}
\usepackage{mathtools}
\usepackage{multirow} 
\usepackage{pgfplots}
\usepackage{physics}
\usepackage[prefix=]{xcolor-material}
\usepackage{tikz}
\renewcommand{\figurename}{Fig.}
\bibpunct[, ]{[}{]}{,}{n}{,}{,}

\makeatletter
\def\NAT@def@citea{\def\@citea{\NAT@separator}}
\makeatother
\begin{document}
\articletype{ARTICLE TEMPLATE}
\title{Soccer Goalkeeper Performance Evaluation: Clustering Approach}
\author{
\name{Mahdi Teimouri \thanks{CONTACT Mahdi Teimouri. Email: mahdiba\_2001@yahoo.com}}
\affil{Department of Statistics, Faculty of Science and Engineering, Gonbad Kavous University, Gonbad Kavous, Iran.}}
\maketitle
\begin{abstract}
The objective of this work was to help the soccer field managers to evaluate the performance of a goalkeeper in saving the penalty kicks. To this end, based on the concept of clustering, four measures were proposed for evaluating the goalkeeper's performance in terms of both the saved kicks and detecting the direction of kicked ball. The well-known measures ignore the goalkeeper's ability in detecting the directional jump, while the forth proposed measure in this work was regarded this fact. The effectiveness of the proposed measures were analyzed and demonstrated by evaluating the performance of goalkeepers participated in four important soccer matches. The results were consistent with the proposed measures. In summary, as well as the known point statistics for evaluating the performance of goalkeeper, the measures were proposed in this work that regard as well the ability of goalkeeper in detecting the direction of kicks are also suggested.
\end{abstract}
\section{Introduction}
A wide range of studies have been devoted to the performance analysis of the soccer players in the literature. These works have considered the effect of several factors such as anthropometric, environmental, physiological, and etc. on the performance of field players regardless of their position \citep{gil2007physiological,wilson2009anxiety,sporis2009fitness,stolen2005physiology,
wong2009relationship,slimani2017anthropometric}. For the goal position, although the activity of goalkeeper is not as much as the field player, but very important in the final result of the match \citep{di2008activity}. For instance, the result of around 28\% (two out of seven) of matches in EURO 2012 has been determined in the knock-out stage \citep{navarro2013mere}). Hence, several attempts have been made in the literature to deal with the factors that affect the performance of goalkeeper, among them for example, we refer to 
\citep{gil2007physiological,ziv2011physical,rebelo2015anthropometric} (for studying the anthropometric situation and physiological performance of goalkeeper), \citep{di2008activity,sainz2008analysis} (for goalkeeper's physical activity and mobility), \citep{navarro2013mere} (for investigating the relation between goalkeeper-independent strategy and penalty kicker target location), \citep{rodriguez2019assessing} (for using the electronic tools to assess the performance of goalkeeper), \citep{alves2010short,perez2023soccer} (for investigating the effect of such factors as sprint, jump, agility, strength, aerobic capacity, mobility, and specific game technique on goalkeeper's performance), \citep{longo2018functional} (for investigating the effect of Italian goalkeeper's activity on team success), and \citep{al2022effect} (for investigating the training program on the goalkeeper's physical abilities). Note that in contrast to the works mentioned above that primarily focus on the factors that affect the performance of a soccer goalkeeper, in this work, we aim to propose some measures to evaluate and then compare the performance of goalkeepers. There are some simple point statistics such as save percentage (SV) defined as
\begin{align}\label{sv}
{\text{SV}}=\frac{{\text{saves}}}{{\text{saves}}+{\text{allowed goals}}},
\end{align}
and the goalkeeper's goals-against average (GAA) defined as
\begin{align}\label{gaa}
{\text{GAA}}=\frac{{\text{allowed goals}} \times 90 }{{\text{total playing time (in minutes)}}},
\end{align}
for evaluating the goalkeeper's performance. It is well known that the soccer goalkeeper needs to be quick in movements, directional changes, and vertical jump \cite{knoop2013evaluation}. Though such measures as (\ref{sv}) and (\ref{gaa}) are useful, but not informative enough for evaluating the movement, directional change, and vertical jump. In fact, the measures in (\ref{sv}) and (\ref{gaa}) focus only the saved and allowed goals, and ignore the goalkeeper's ability in directional change or jump.
\par In this work, we propose four new measures that can evaluate the performance of a goalkeeper during the penalty kicks that regard the goalkeeper's ability in directional change or jump. The first three measures are constructed based on the concept of clustering that is an unsupervised machine learning tool and the fourth one is a modification of the SV measure given in (\ref{sv}).
\subsection{Material and method}
\subsection{Material}
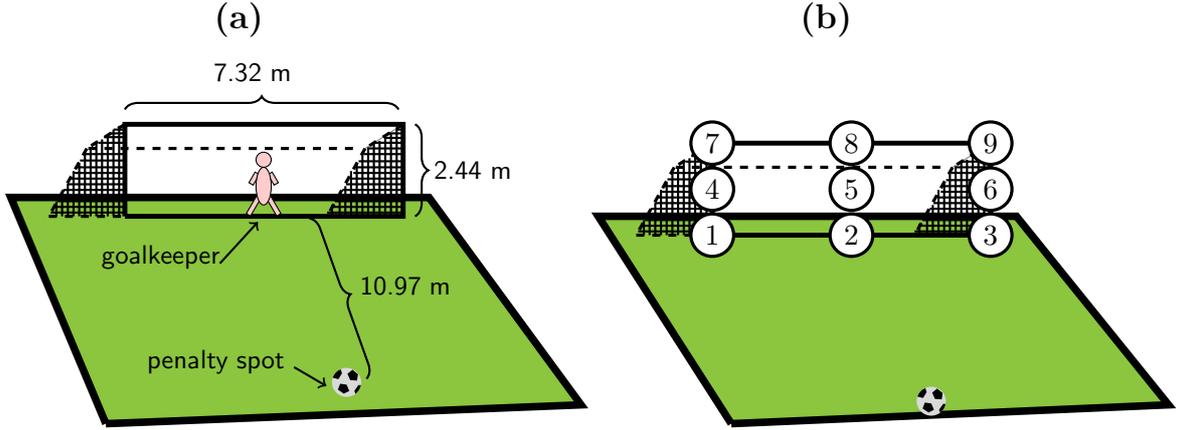
\begin{figure}[!h]
  \centering
\begin{tikzpicture}[scale=.5]
\draw[line width=1mm,fill=LimeGreen] (-.5,-5.5) -- (12,-5) -- (8,.5) -- (-3,.50)--(-.5,-5.5) node{};
\draw[line width=.65mm][solid] (0,0)--(7.32,0)--(7.32,2.44)--(0,2.44)--(0,0);
\draw[very thick][fill=gray][dashed] (7.32,2.44)--(6.32,1.8);
\draw[very thick][dashed] (6.32,1.8)--(5.32,0);
\draw[very thick][dashed] (6,1.8)--(-1,1.8);
\draw[very thick][dashed] (0,2.44)--(-1,1.8);
\draw[very thick][dashed] (-1,1.8)--(-2,0);
\draw[very thick][dashed] (-2,0)--(0,0);
\draw[-,thick] (-1.95,.10)--(0,.10);
\draw[-,thick] (-1.85,.25)--(0,.250);
\draw[-,thick] (-1.78,.4)--(0,.4);
\draw[-,thick] (-1.65,.55)--(0,.55);
\draw[-,thick] (-1.62,.7)--(0,.7);
\draw[-,thick] (-1.52,.85)--(0,.85);
\draw[-,thick] (-1.46,1)--(0,1);
\draw[-,thick] (-1.35,1.15)--(0,1.15);
\draw[-,thick] (-1.28,1.3)--(0,1.3);
\draw[-,thick] (-1.18,1.48)--(0,1.48);
\draw[-,thick] (-1.1,1.65)--(0,1.65);
\draw[-,thick] (-.7,1.95)--(0,1.95);
\draw[-,thick] (-.55,2.1)--(0,2.1);
\draw[-,thick] (-.4,2.2)--(0,2.2);
\draw[-,thick] (-.3,2.3)--(0,2.3);
\draw[-,thick] (5.32,.10)--(7.32,.10);
\draw[-,thick] (5.42,.25)--(7.32,.250);
\draw[-,thick] (5.52,.4)--(7.32,.4);
\draw[-,thick] (5.62,.55)--(7.32,.55);%
\draw[-,thick] (5.72,.7)--(7.32,.7);
\draw[-,thick] (5.82,.85)--(7.32,.85);
\draw[-,thick] (5.92,1)--(7.32,1);
\draw[-,thick] (6.01,1.15)--(7.32,1.15);
\draw[-,thick] (6.10,1.3)--(7.32,1.3);
\draw[-,thick] (6.15,1.46)--(7.32,1.46);
\draw[-,thick] (6.28,1.65)--(7.32,1.65);%
\draw[-,thick] (6.41,1.8)--(7.32,1.8);
\draw[-,thick] (6.53,1.95)--(7.32,1.95);
\draw[-,thick] (6.85,2.08)--(7.32,2.08);
\draw[-,thick] (6.92,2.2)--(7.32,2.2);
\draw[-,thick] (7.15,2.3)--(7.32,2.3);
\draw[-,thick] (7.15,2.34)--(7.15,0);
\draw[-,thick] (7.0,2.2)--(7.0,0);
\draw[-,thick] (6.85,2.13)--(6.85,0);
\draw[-,thick] (6.7,2)--(6.7,0);
\draw[-,thick] (6.55,1.94)--(6.55,0);
\draw[-,thick] (6.4,1.83)--(6.4,0);
\draw[-,thick] (6.25,1.65)--(6.25,0);
\draw[-,thick] (6.1,1.4)--(6.1,0);
\draw[-,thick] (5.95,1.2)--(5.95,0);
\draw[-,thick] (5.8,.8)--(5.8,0);
\draw[-,thick] (5.65,.6)--(5.65,0);
\draw[-,thick] (5.45,.3)--(5.45,0);
\draw[-,thick] (-.15,2.4)--(-.15,0);
\draw[-,thick] (-.3,2.3)--(-.3,0);
\draw[-,thick] (-.45,2.13)--(-.45,0);
\draw[-,thick] (-.6,2.04)--(-.6,0);
\draw[-,thick] (-.75,1.94)--(-.75,0);
\draw[-,thick] (-.9,1.83)--(-.9,0); %
\draw[-,thick] (-1.05,1.65)--(-1.05,0);
\draw[-,thick] (-1.2,1.4)--(-1.2,0);
\draw[-,thick] (-1.35,1.2)--(-1.35,0);
\draw[-,thick] (-1.5,.85)--(-1.5,0);
\draw[-,thick] (-1.65,.56)--(-1.65,0);
\draw[-,thick] (-1.8,.39)--(-1.8,0);
\begin{scope}[scale=1.2,shift={(3.66,-4)}]
\fill[gray!30!white] (1.2,0.33) circle (0.32);
\clip (1.2,0.33) circle (0.32);
\fill[black] (1.06,0.30)--(1.01,0.17)--(1.14,0.08)--(1.26,0.14)--(1.20,0.28) -- cycle (1.37,0.14) -- (1.46,0.27) -- (1.59,0.27) -- (1.41,0.04) -- cycle (1.28,0.38) -- (1.22,0.52) -- (1.33,0.61) -- (1.45,0.51) -- (1.43,0.37) -- cycle (0.87,0.44) -- (1.02,0.40) -- (1.10,0.53) -- (1.07,0.62) -- (0.94,0.57) -- cycle;
\end{scope}
\draw [thick,decorate,decoration={brace,amplitude=5pt,mirror,raise=4ex}]  (5.1,-4.75) -- (3.6,-.5) node[midway,right,font=\small\sffamily,xshift=2em,yshift=1em]{10.97 m};
\draw [thick,decorate,decoration={brace,amplitude=5pt,mirror,raise=4ex}]  (6.3,0) -- (6.3,2.44) node[midway,right,font=\small\sffamily,xshift=2em,yshift=0em]{2.44 m};
\draw [thick,decorate,decoration={brace,amplitude=5pt,raise=4ex}]  (0,1.5) -- (7.2,1.5) node[midway,right,font=\small\sffamily,xshift=-2em,yshift=3em]{7.32 m};
\draw[->,thick] (4.45,-4)  -- node [thick,midway,left,font=\small\sffamily,xshift=-.50em,yshift=.5em] {penalty spot} (5.3,-4.5);
\draw[->,thick] (2.5,-1.25)  -- node [thick,midway,left,font=\small\sffamily,xshift=-.30em,yshift=-.5em] {goalkeeper} (3.5,-.15);
  \draw[fill=red!20](3.65,1.5) circle (.20) node [black,yshift=-1.5cm] {};
\begin{scope}[scale=.5,transform canvas={xshift=1.7cm,yshift=.15cm}]
\node [rotate=-30,fill=red!20,transform shape,draw,rectangle,minimum width=.05cm,minimum height=1.2cm] {};
\end{scope}
\begin{scope}[scale=.5,transform canvas={xshift=1.95cm,yshift=.150cm}]
\node [rotate=30,fill=red!20,transform shape,draw,rectangle,minimum width=.05cm,minimum height=1.2cm] {};
\end{scope}
\begin{scope}[scale=.5,transform canvas={xshift=1.7cm,yshift=.5cm}]
\node [rotate=-30,fill=red!20,transform shape,draw,rectangle,minimum width=.05cm,minimum height=.65cm] {};
\end{scope}
\begin{scope}[scale=.5,transform canvas={xshift=1.95cm,yshift=.5cm}]
\node [rotate=30,fill=red!20,transform shape,draw,rectangle,minimum width=.05cm,minimum height=.65cm] {};
\end{scope}
\draw[rotate=0,fill=red!20] (3.65, .8) ellipse (0.2cm and .5cm);
\node[above,font=\large\bfseries] at (3,4.5) {(a)};
\end{tikzpicture}
\begin{tikzpicture}[scale=.5]
\draw[line width=1mm,fill=LimeGreen] (.5,-5) -- (12,-4.5) -- (8,.5) -- (-3,.50)--(.5,-5) node{};
\draw[line width=.65mm][solid] (0,0)--(7.32,0)--(7.32,2.44)--(0,2.44)--(0,0);
\draw[very thick][fill=gray][dashed] (7.32,2.44)--(6.32,1.8);
\draw[very thick][dashed] (6.32,1.8)--(5.32,0);
\draw[very thick][dashed] (6,1.8)--(-1,1.8);
\draw[very thick][dashed] (0,2.44)--(-1,1.8);
\draw[very thick][dashed] (-1,1.8)--(-2,0);
\draw[very thick][dashed] (-2,0)--(0,0);
\draw[-,thick] (-1.95,.10)--(0,.10);
\draw[-,thick] (-1.85,.25)--(0,.250);
\draw[-,thick] (-1.78,.4)--(0,.4);
\draw[-,thick] (-1.65,.55)--(0,.55);
\draw[-,thick] (-1.62,.7)--(0,.7);
\draw[-,thick] (-1.52,.85)--(0,.85);
\draw[-,thick] (-1.46,1)--(0,1);
\draw[-,thick] (-1.35,1.15)--(0,1.15);
\draw[-,thick] (-1.28,1.3)--(0,1.3);
\draw[-,thick] (-1.18,1.48)--(0,1.48);
\draw[-,thick] (-1.1,1.65)--(0,1.65);
\draw[-,thick] (-.7,1.95)--(0,1.95);
\draw[-,thick] (-.55,2.1)--(0,2.1);
\draw[-,thick] (-.4,2.2)--(0,2.2);
\draw[-,thick] (-.3,2.3)--(0,2.3);
\draw[-,thick] (5.32,.10)--(7.32,.10);
\draw[-,thick] (5.42,.25)--(7.32,.250);
\draw[-,thick] (5.52,.4)--(7.32,.4);
\draw[-,thick] (5.62,.55)--(7.32,.55);%
\draw[-,thick] (5.72,.7)--(7.32,.7);
\draw[-,thick] (5.82,.85)--(7.32,.85);
\draw[-,thick] (5.92,1)--(7.32,1);
\draw[-,thick] (6.01,1.15)--(7.32,1.15);
\draw[-,thick] (6.10,1.3)--(7.32,1.3);
\draw[-,thick] (6.15,1.46)--(7.32,1.46);
\draw[-,thick] (6.28,1.65)--(7.32,1.65);%
\draw[-,thick] (6.41,1.8)--(7.32,1.8);
\draw[-,thick] (6.53,1.95)--(7.32,1.95);
\draw[-,thick] (6.85,2.08)--(7.32,2.08);
\draw[-,thick] (6.92,2.2)--(7.32,2.2);
\draw[-,thick] (7.15,2.3)--(7.32,2.3);
\draw[-,thick] (7.15,2.34)--(7.15,0);
\draw[-,thick] (7.0,2.2)--(7.0,0);
\draw[-,thick] (6.85,2.13)--(6.85,0);
\draw[-,thick] (6.7,2)--(6.7,0);
\draw[-,thick] (6.55,1.94)--(6.55,0);
\draw[-,thick] (6.4,1.83)--(6.4,0);
\draw[-,thick] (6.25,1.65)--(6.25,0);
\draw[-,thick] (6.1,1.4)--(6.1,0);
\draw[-,thick] (5.95,1.2)--(5.95,0);
\draw[-,thick] (5.8,.8)--(5.8,0);
\draw[-,thick] (5.65,.6)--(5.65,0);
\draw[-,thick] (5.45,.3)--(5.45,0);
\draw[-,thick] (-.15,2.4)--(-.15,0);
\draw[-,thick] (-.3,2.3)--(-.3,0);
\draw[-,thick] (-.45,2.13)--(-.45,0);
\draw[-,thick] (-.6,2.04)--(-.6,0);
\draw[-,thick] (-.75,1.94)--(-.75,0);
\draw[-,thick] (-.9,1.83)--(-.9,0); %
\draw[-,thick] (-1.05,1.65)--(-1.05,0);
\draw[-,thick] (-1.2,1.4)--(-1.2,0);
\draw[-,thick] (-1.35,1.2)--(-1.35,0);
\draw[-,thick] (-1.5,.85)--(-1.5,0);
\draw[-,thick] (-1.65,.56)--(-1.65,0);
\draw[-,thick] (-1.8,.39)--(-1.8,0);
\begin{scope}[scale=1.2,shift={(3.6,-4)}]
      \fill[gray!30!white] (1.2,0.33) circle (0.32);
      \clip (1.2,0.33) circle (0.32);
      \fill[black] (1.06,0.30) -- (1.01,0.17) -- (1.14,0.08) -- (1.26,0.14) -- (1.20,0.28) -- cycle (1.37,0.14) -- (1.46,0.27) -- (1.59,0.27) -- (1.41,0.04) -- cycle (1.28,0.38) -- (1.22,0.52) -- (1.33,0.61) -- (1.45,0.51) -- (1.43,0.37) -- cycle (0.87,0.44) -- (1.02,0.40) -- (1.10,0.53) -- (1.07,0.62) -- (0.94,0.57) -- cycle;
\end{scope}
\draw[very thick](0,0)[fill=white] circle (16pt) node{1};
\draw[very thick](0,1.22)[fill=white] circle (16pt) node{4};
\draw[very thick](0,2.44)[fill=white] circle (16pt) node{7};
\draw[very thick](3.66,0)[fill=white] circle (16pt) node{2};
\draw[very thick](3.66,1.22)[fill=white] circle (16pt) node{5};
\draw[very thick](3.66,2.44)[fill=white] circle (16pt) node{8};
\draw[very thick](7.32,0)[fill=white] circle (16pt) node{3};
\draw[very thick](7.32,1.22)[fill=white] circle (16pt) node{6};
\draw[very thick](7.32,2.44)[fill=white] circle (16pt) node{9};
\node[above,font=\large\bfseries] at (3,5) {(b)};
\end{tikzpicture}
\caption{(a): Specifications of football pitch around the penalty point. (b) Nine clusters numbered with $1,2,\cdots,9$ that are assumed to be the target of penalty shootouts.}
\label{fig1}
\end{figure}

Suppose a goal soccer is given a penalty kick at penalty point and the goalkeeper stands in the middle of goal line as shown in Figure \ref{fig1} (a). The penalty kicker kicks the ball into one of nine zones (clusters) as shown in Figure \ref{fig1} (b). It is reasonable to assume that the goalkeeper has no prior information about the direction of kicked ball to the goal.
\subsection{Method}\label{method}
Suppose there are two clustering approaches $M_{1}$ and $M_{2}$ for partitioning a sample of $n$ observations $S=\bigl\{o_1,o_2,\cdots,o_n\bigr\}$ into $r$ and $c$ clusters shown by $M_{1}=\bigl\{C_{11},C_{12},\cdots,C_{1r}\bigr\}$ and $M_{2}=\bigl\{C_{21},C_{22},\cdots,C_{2c}\bigr\}$, respectively. Table \ref{tab1} shows a contingency table in which $n_{ij}=\big \vert C_{1i} \cap C_{2j}\big \vert$ (for $i=1,\cdots,r$ and $j=1,\cdots,c$) represents the number of common observations between partitions suggested by clustering approaches $M_{1}$ and $M_{2}$. The Rand index \citep{rand1971objective} (hereafter denoted as RI) is a well-known tool for measuring the similarity between two clustering approaches. Suppose quantities $a$, $b$, $c$, and $d$ are defined as follows.
\begin{itemize}
\item ${a}$ is the number of paired samples in $S$ that are in the same cluster within $M_{1}$ and the same cluster within $M_{2}$.
\item ${b}$ is the number of paired samples in $S$ that are in the different clusters within $M_{1}$ and the different clusters within $M_{2}$.
\item ${c}$ is the number of paired samples in $S$ that are in the same cluster within $M_{1}$ and the different clusters within $M_{2}$.
\item ${d}$ is the number of paired samples in $S$ that are in different clusters within $M_{1}$ and the same cluster within $M_{2}$.
\end{itemize}
Given a set of observations $S$, it is known that there are $n(n-1)/2={a}+{b}+{c}+{d}$ possibilities for drawing paired samples from $S$. The RI is then given by
\begin{align}\label{RI}
{\text{RI}}=\frac{{a}+{b}}{{a}+{b}+{c}+{d}}=2\frac{{a}+{b}}{n(n-1)}.
\end{align}
Obviously $0 \leq {\text{RI}}\leq 1$ and we have ${\text{RI}}=0$, if there is no similarity between two clustering approaches while the upper boundary value of RI reflects the complete similarity between two clustering approaches. This means that larger values of RI indicate more similarities between two clustering approaches.
\begin{table}[!h]
\caption{Contingency table for clustering models $M_1$ and $M_2$.}
\centering
{\begin{tabular}{cccccc}
\hline
\hline
\multirow{1}{*}{}&&\multicolumn{4}{c}{$M_{2}$}\\
\cline{3-6}
    & &$C_{21}$&$C_{22}$&$\cdots$&$C_{2c}$\\
\hline
 \multirow{4}{*}{$M_{1}$}  & $C_{11}$                  &$n_{11}$    & $n_{12}$ &$\cdots$ & $n_{1c}$\\                                                                                             
\cline{2-6}                                                                                     
                          				  & $C_{12}$                  &  $n_{21}$  &  $n_{22}$& $\cdots$& $n_{2c}$\\                                                                                                                                                                                 \cline{2-6}                                                                                     
                          				  & $\vdots $                  &  $\vdots$   & $\vdots$  & $\ddots$&  $\vdots$\\                                                                                                                                                                                                                                                                        
\cline{2-6}                                                                                     
                          				  & $C_{1r}$                  &  $n_{r1}$  &  $n_{r2}$ & $\cdots$& $n_{rc}$\\                                                                                              
\hline
\hline
\end{tabular}}
\label{tab1}
\end{table}
The logic behind the definition of RI motivated us to construct a method for evaluating and comparing the performance of goalkeepers in detecting direction of kicked ball to the goal. Without loss of generality, we suppose the goal zone is partitioned into nine areas as shown in Figure \ref{fig1} (b). The centers of clusters are shown by Figure \ref{fig2} (a) are $\mu_{1},\mu_{2},\cdots,\mu_{9}$. The $x$-coordinate and $y$-coordinate correspond to centers of clusters are $\{0, 3.66, 7.32\}$ and $\{0, 1.22, 2.44\}$, respectively. As it is seen, the centers are equidistant in both coordinates. To describe our proposed method, suppose that goal soccer shown in Figure \ref{fig3} (a) kicked the ball at first zone (cluster). This means that the true cluster for first kicked ball (observation) is the first zone with center $\mu_{1}$. As it is seen, the goalkeeper assigns it incorrectly to the third zone with center $\mu_{3}$. Suppose the true clusters under name $M_{true}$ is determined by the goal kicker and the goal keeper is also detecting clusters under name $M_{1}$, hence for the first observation, we have $M_{true}=\{1\}$ and $M_{1}=\{3\}$. Let the second kick is shown by Figure \ref{fig3}(b). In this case, we have $M_{true}=\{7\}$ and $M_{1}=\{7\}$. Based on two observations, we can write $M_{true}=\{1,7\}$ and $M_{1}=\{3,7\}$. This process can be repeated even for a long run. It is worthwhile to mention that when the goalkeeper detects the cluster correctly, then she/he may save, touch, or not touch the kicked ball.
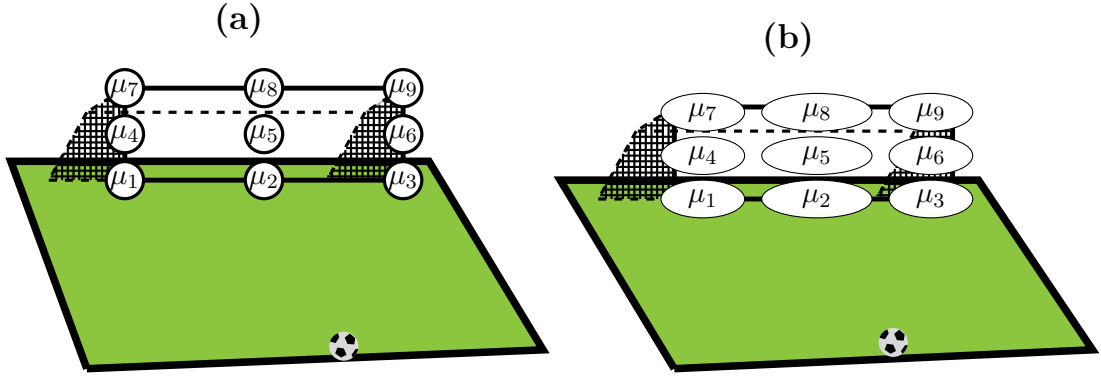
\begin{figure}[!h]
  \centering
\begin{tikzpicture}[scale=.5]
\draw[line width=1mm,fill=LimeGreen] (-1,-5) -- (11,-4.5) -- (8,.5) -- (-3,.50)--(-1,-5) node{};
\draw[line width=.65mm][solid] (0,0)--(7.32,0)--(7.32,2.44)--(0,2.44)--(0,0);
\draw[very thick][fill=gray][dashed] (7.32,2.44)--(6.32,1.8);
\draw[very thick][dashed] (6.32,1.8)--(5.32,0);
\draw[very thick][dashed] (6,1.8)--(-1,1.8);
\draw[very thick][dashed] (0,2.44)--(-1,1.8);
\draw[very thick][dashed] (-1,1.8)--(-2,0);
\draw[very thick][dashed] (-2,0)--(0,0);
\draw[-,thick] (-1.95,.10)--(0,.10);
\draw[-,thick] (-1.85,.25)--(0,.250);
\draw[-,thick] (-1.78,.4)--(0,.4);
\draw[-,thick] (-1.65,.55)--(0,.55);
\draw[-,thick] (-1.62,.7)--(0,.7);
\draw[-,thick] (-1.52,.85)--(0,.85);
\draw[-,thick] (-1.46,1)--(0,1);
\draw[-,thick] (-1.35,1.15)--(0,1.15);
\draw[-,thick] (-1.28,1.3)--(0,1.3);
\draw[-,thick] (-1.18,1.48)--(0,1.48);
\draw[-,thick] (-1.1,1.65)--(0,1.65);
\draw[-,thick] (-.7,1.95)--(0,1.95);
\draw[-,thick] (-.55,2.1)--(0,2.1);
\draw[-,thick] (-.4,2.2)--(0,2.2);
\draw[-,thick] (-.3,2.3)--(0,2.3);
\draw[-,thick] (5.32,.10)--(7.32,.10);
\draw[-,thick] (5.42,.25)--(7.32,.250);
\draw[-,thick] (5.52,.4)--(7.32,.4);
\draw[-,thick] (5.62,.55)--(7.32,.55);%
\draw[-,thick] (5.72,.7)--(7.32,.7);
\draw[-,thick] (5.82,.85)--(7.32,.85);
\draw[-,thick] (5.92,1)--(7.32,1);
\draw[-,thick] (6.01,1.15)--(7.32,1.15);
\draw[-,thick] (6.10,1.3)--(7.32,1.3);
\draw[-,thick] (6.15,1.46)--(7.32,1.46);
\draw[-,thick] (6.28,1.65)--(7.32,1.65);%
\draw[-,thick] (6.41,1.8)--(7.32,1.8);
\draw[-,thick] (6.53,1.95)--(7.32,1.95);
\draw[-,thick] (6.85,2.08)--(7.32,2.08);
\draw[-,thick] (6.92,2.2)--(7.32,2.2);
\draw[-,thick] (7.15,2.3)--(7.32,2.3);
\draw[-,thick] (7.15,2.34)--(7.15,0);
\draw[-,thick] (7.0,2.2)--(7.0,0);
\draw[-,thick] (6.85,2.13)--(6.85,0);
\draw[-,thick] (6.7,2)--(6.7,0);
\draw[-,thick] (6.55,1.94)--(6.55,0);
\draw[-,thick] (6.4,1.83)--(6.4,0);
\draw[-,thick] (6.25,1.65)--(6.25,0);
\draw[-,thick] (6.1,1.4)--(6.1,0);
\draw[-,thick] (5.95,1.2)--(5.95,0);
\draw[-,thick] (5.8,.8)--(5.8,0);
\draw[-,thick] (5.65,.6)--(5.65,0);
\draw[-,thick] (5.45,.3)--(5.45,0);
\draw[-,thick] (-.15,2.4)--(-.15,0);
\draw[-,thick] (-.3,2.3)--(-.3,0);
\draw[-,thick] (-.45,2.13)--(-.45,0);
\draw[-,thick] (-.6,2.04)--(-.6,0);
\draw[-,thick] (-.75,1.94)--(-.75,0);
\draw[-,thick] (-.9,1.83)--(-.9,0); %
\draw[-,thick] (-1.05,1.65)--(-1.05,0);
\draw[-,thick] (-1.2,1.4)--(-1.2,0);
\draw[-,thick] (-1.35,1.2)--(-1.35,0);
\draw[-,thick] (-1.5,.85)--(-1.5,0);
\draw[-,thick] (-1.65,.56)--(-1.65,0);
\draw[-,thick] (-1.8,.39)--(-1.8,0);
\begin{scope}[scale=1.2,shift={(3.6,-4)}]
      \fill[gray!30!white] (1.2,0.33) circle (0.32);
      \clip (1.2,0.33) circle (0.32);
      \fill[black] (1.06,0.30) -- (1.01,0.17) -- (1.14,0.08) -- (1.26,0.14) -- (1.20,0.28) -- cycle (1.37,0.14) -- (1.46,0.27) -- (1.59,0.27) -- (1.41,0.04) -- cycle (1.28,0.38) -- (1.22,0.52) -- (1.33,0.61) -- (1.45,0.51) -- (1.43,0.37) -- cycle (0.87,0.44) -- (1.02,0.40) -- (1.10,0.53) -- (1.07,0.62) -- (0.94,0.57) -- cycle;
\end{scope}
\draw[very thick](0,0)[fill=white] circle (14pt) node{$\mu_{1}$};
\draw[very thick](0,1.22)[fill=white] circle (14pt) node{$\mu_{4}$};
\draw[very thick](0,2.44)[fill=white] circle (14pt) node{$\mu_{7}$};
\draw[very thick](3.66,0)[fill=white] circle (14pt) node{$\mu_{2}$};
\draw[very thick](3.66,1.22)[fill=white] circle (14pt) node{$\mu_{5}$};
\draw[very thick](3.66,2.44)[fill=white] circle (14pt) node{$\mu_{8}$};
\draw[very thick](7.32,0)[fill=white] circle (14pt) node{$\mu_{3}$};
\draw[very thick](7.32,1.22)[fill=white] circle (14pt) node{$\mu_{6}$};
\draw[very thick](7.32,2.44)[fill=white] circle (14pt) node{$\mu_{9}$};
\node[above,font=\large\bfseries] at (3,3.5) {(a)};
\end{tikzpicture}
\begin{tikzpicture}[scale=.5]
\draw[line width=1mm,fill=LimeGreen] (0,-4.5) -- (11,-4) -- (8,.5) -- (-3,.50)--(0,-4.5) node{};
\draw[line width=.65mm][solid] (0,0)--(7.32,0)--(7.32,2.44)--(0,2.44)--(0,0);
\draw[very thick][fill=gray][dashed] (7.32,2.44)--(6.32,1.8);
\draw[very thick][dashed] (6.32,1.8)--(5.32,0);
\draw[very thick][dashed] (6,1.8)--(-1,1.8);
\draw[very thick][dashed] (0,2.44)--(-1,1.8);
\draw[very thick][dashed] (-1,1.8)--(-2,0);
\draw[very thick][dashed] (-2,0)--(0,0);
\draw[-,thick] (-1.95,.10)--(0,.10);
\draw[-,thick] (-1.85,.25)--(0,.250);
\draw[-,thick] (-1.78,.4)--(0,.4);
\draw[-,thick] (-1.65,.55)--(0,.55);
\draw[-,thick] (-1.62,.7)--(0,.7);
\draw[-,thick] (-1.52,.85)--(0,.85);
\draw[-,thick] (-1.46,1)--(0,1);
\draw[-,thick] (-1.35,1.15)--(0,1.15);
\draw[-,thick] (-1.28,1.3)--(0,1.3);
\draw[-,thick] (-1.18,1.48)--(0,1.48);
\draw[-,thick] (-1.1,1.65)--(0,1.65);
\draw[-,thick] (-.7,1.95)--(0,1.95);
\draw[-,thick] (-.55,2.1)--(0,2.1);
\draw[-,thick] (-.4,2.2)--(0,2.2);
\draw[-,thick] (-.3,2.3)--(0,2.3);
\draw[-,thick] (5.32,.10)--(7.32,.10);
\draw[-,thick] (5.42,.25)--(7.32,.250);
\draw[-,thick] (5.52,.4)--(7.32,.4);
\draw[-,thick] (5.62,.55)--(7.32,.55);%
\draw[-,thick] (5.72,.7)--(7.32,.7);
\draw[-,thick] (5.82,.85)--(7.32,.85);
\draw[-,thick] (5.92,1)--(7.32,1);
\draw[-,thick] (6.01,1.15)--(7.32,1.15);
\draw[-,thick] (6.10,1.3)--(7.32,1.3);
\draw[-,thick] (6.15,1.46)--(7.32,1.46);
\draw[-,thick] (6.28,1.65)--(7.32,1.65);%
\draw[-,thick] (6.41,1.8)--(7.32,1.8);
\draw[-,thick] (6.53,1.95)--(7.32,1.95);
\draw[-,thick] (6.85,2.08)--(7.32,2.08);
\draw[-,thick] (6.92,2.2)--(7.32,2.2);
\draw[-,thick] (7.15,2.3)--(7.32,2.3);
\draw[-,thick] (7.15,2.34)--(7.15,0);
\draw[-,thick] (7.0,2.2)--(7.0,0);
\draw[-,thick] (6.85,2.13)--(6.85,0);
\draw[-,thick] (6.7,2)--(6.7,0);
\draw[-,thick] (6.55,1.94)--(6.55,0);
\draw[-,thick] (6.4,1.83)--(6.4,0);
\draw[-,thick] (6.25,1.65)--(6.25,0);
\draw[-,thick] (6.1,1.4)--(6.1,0);
\draw[-,thick] (5.95,1.2)--(5.95,0);
\draw[-,thick] (5.8,.8)--(5.8,0);
\draw[-,thick] (5.65,.6)--(5.65,0);
\draw[-,thick] (5.45,.3)--(5.45,0);
\draw[-,thick] (-.15,2.4)--(-.15,0);
\draw[-,thick] (-.3,2.3)--(-.3,0);
\draw[-,thick] (-.45,2.13)--(-.45,0);
\draw[-,thick] (-.6,2.04)--(-.6,0);
\draw[-,thick] (-.75,1.94)--(-.75,0);
\draw[-,thick] (-.9,1.83)--(-.9,0); %
\draw[-,thick] (-1.05,1.65)--(-1.05,0);
\draw[-,thick] (-1.2,1.4)--(-1.2,0);
\draw[-,thick] (-1.35,1.2)--(-1.35,0);
\draw[-,thick] (-1.5,.85)--(-1.5,0);
\draw[-,thick] (-1.65,.56)--(-1.65,0);
\draw[-,thick] (-1.8,.39)--(-1.8,0);
\begin{scope}[scale=1.2,shift={(3.6,-3.5)}]
      \fill[gray!30!white] (1.2,0.33) circle (0.32);
      \clip (1.2,0.33) circle (0.32);
      \fill[black] (1.06,0.30) -- (1.01,0.17) -- (1.14,0.08) -- (1.26,0.14) -- (1.20,0.28) -- cycle (1.37,0.14) -- (1.46,0.27) -- (1.59,0.27) -- (1.41,0.04) -- cycle (1.28,0.38) -- (1.22,0.52) -- (1.33,0.61) -- (1.45,0.51) -- (1.43,0.37) -- cycle (0.87,0.44) -- (1.02,0.40) -- (1.10,0.53) -- (1.07,0.62) -- (0.94,0.57) -- cycle;
\end{scope}
\draw[rotate=90,fill=white] (0,-.75) ellipse (0.5cm and 1.1cm) node{$\mu_{1}$};
\draw[rotate=90,fill=white] (0,-3.75) ellipse (0.5cm and 1.45cm) node{$\mu_{2}$};
\draw[rotate=90,fill=white] (0,-6.75) ellipse (0.5cm and 1.1cm) node{$\mu_{3}$};
\draw[rotate=90,fill=white] (1.15,-.75) ellipse (0.5cm and 1.1cm) node{$\mu_{4}$};
\draw[rotate=90,fill=white] (1.15,-3.75) ellipse (0.5cm and 1.45cm) node{$\mu_{5}$};
\draw[rotate=90,fill=white] (1.15,-6.75) ellipse (0.5cm and 1.1cm) node{$\mu_{6}$};
\draw[rotate=90,fill=white] (2.3,-.75) ellipse (0.5cm and 1.1cm) node{$\mu_{7}$};
\draw[rotate=90,fill=white] (2.3,-3.75) ellipse (0.5cm and 1.45cm) node{$\mu_{8}$};
\draw[rotate=90,fill=white] (2.3,-6.75) ellipse (0.5cm and 1.1cm) node{$\mu_{9}$};
\node[above,font=\large\bfseries] at (3,3.5) {(b)};
\end{tikzpicture}
\caption{(a): Centers of nine circular clusters. (b) Centers of nine elliptical clusters.}
\label{fig2}
\end{figure}
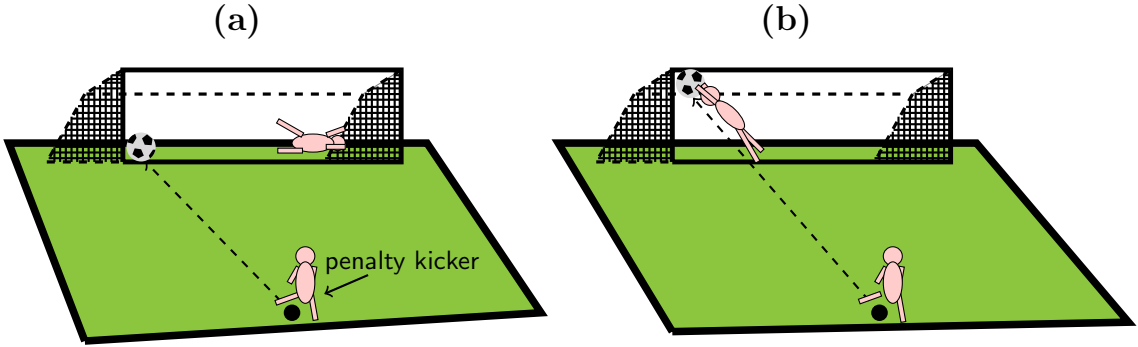
\begin{figure}[!h]
  \centering
\begin{tikzpicture}[scale=.5]
\draw[line width=1mm,fill=LimeGreen] (-1,-4.75) -- (11,-4) -- (8,.5) -- (-3,.50)--(-1,-4.75) node{};
\draw[line width=.65mm][solid] (0,0)--(7.32,0)--(7.32,2.44)--(0,2.44)--(0,0);
\draw[very thick][fill=gray][dashed] (7.32,2.44)--(6.32,1.8);
\draw[very thick][dashed] (6.32,1.8)--(5.32,0);
\draw[very thick][dashed] (6,1.8)--(-1,1.8);
\draw[very thick][dashed] (0,2.44)--(-1,1.8);
\draw[very thick][dashed] (-1,1.8)--(-2,0);
\draw[very thick][dashed] (-2,0)--(0,0);
\draw[-,thick] (-1.95,.10)--(0,.10);
\draw[-,thick] (-1.85,.25)--(0,.250);
\draw[-,thick] (-1.78,.4)--(0,.4);
\draw[-,thick] (-1.65,.55)--(0,.55);
\draw[-,thick] (-1.62,.7)--(0,.7);
\draw[-,thick] (-1.52,.85)--(0,.85);
\draw[-,thick] (-1.46,1)--(0,1);
\draw[-,thick] (-1.35,1.15)--(0,1.15);
\draw[-,thick] (-1.28,1.3)--(0,1.3);
\draw[-,thick] (-1.18,1.48)--(0,1.48);
\draw[-,thick] (-1.1,1.65)--(0,1.65);
\draw[-,thick] (-.7,1.95)--(0,1.95);
\draw[-,thick] (-.55,2.1)--(0,2.1);
\draw[-,thick] (-.4,2.2)--(0,2.2);
\draw[-,thick] (-.3,2.3)--(0,2.3);
\draw[-,thick] (5.32,.10)--(7.32,.10);
\draw[-,thick] (5.42,.25)--(7.32,.250);
\draw[-,thick] (5.52,.4)--(7.32,.4);
\draw[-,thick] (5.62,.55)--(7.32,.55);%
\draw[-,thick] (5.72,.7)--(7.32,.7);
\draw[-,thick] (5.82,.85)--(7.32,.85);
\draw[-,thick] (5.92,1)--(7.32,1);
\draw[-,thick] (6.01,1.15)--(7.32,1.15);
\draw[-,thick] (6.10,1.3)--(7.32,1.3);
\draw[-,thick] (6.15,1.46)--(7.32,1.46);
\draw[-,thick] (6.28,1.65)--(7.32,1.65);%
\draw[-,thick] (6.41,1.8)--(7.32,1.8);
\draw[-,thick] (6.53,1.95)--(7.32,1.95);
\draw[-,thick] (6.85,2.08)--(7.32,2.08);
\draw[-,thick] (6.92,2.2)--(7.32,2.2);
\draw[-,thick] (7.15,2.3)--(7.32,2.3);
\draw[-,thick] (7.15,2.34)--(7.15,0);
\draw[-,thick] (7.0,2.2)--(7.0,0);
\draw[-,thick] (6.85,2.13)--(6.85,0);
\draw[-,thick] (6.7,2)--(6.7,0);
\draw[-,thick] (6.55,1.94)--(6.55,0);
\draw[-,thick] (6.4,1.83)--(6.4,0);
\draw[-,thick] (6.25,1.65)--(6.25,0);
\draw[-,thick] (6.1,1.4)--(6.1,0);
\draw[-,thick] (5.95,1.2)--(5.95,0);
\draw[-,thick] (5.8,.8)--(5.8,0);
\draw[-,thick] (5.65,.6)--(5.65,0);
\draw[-,thick] (5.45,.3)--(5.45,0);
\draw[-,thick] (-.15,2.4)--(-.15,0);
\draw[-,thick] (-.3,2.3)--(-.3,0);
\draw[-,thick] (-.45,2.13)--(-.45,0);
\draw[-,thick] (-.6,2.04)--(-.6,0);
\draw[-,thick] (-.75,1.94)--(-.75,0);
\draw[-,thick] (-.9,1.83)--(-.9,0); %
\draw[-,thick] (-1.05,1.65)--(-1.05,0);
\draw[-,thick] (-1.2,1.4)--(-1.2,0);
\draw[-,thick] (-1.35,1.2)--(-1.35,0);
\draw[-,thick] (-1.5,.85)--(-1.5,0);
\draw[-,thick] (-1.65,.56)--(-1.65,0);
\draw[-,thick] (-1.8,.39)--(-1.8,0);
\begin{scope}[scale=1.2,shift={(-0.8,0)}]
      \fill[gray!30!white] (1.2,0.33) circle (0.32);
      \clip (1.2,0.33) circle (0.32);
      \fill[black] (1.06,0.30) -- (1.01,0.17) -- (1.14,0.08) -- (1.26,0.14) -- (1.20,0.28) -- cycle (1.37,0.14) -- (1.46,0.27) -- (1.59,0.27) -- (1.41,0.04) -- cycle (1.28,0.38) -- (1.22,0.52) -- (1.33,0.61) -- (1.45,0.51) -- (1.43,0.37) -- cycle (0.87,0.44) -- (1.02,0.40) -- (1.10,0.53) -- (1.07,0.62) -- (0.94,0.57) -- cycle;
\end{scope}
\draw[very thick](4.45,-4)[fill=black] circle (5pt) node{};
\draw[->,thick,dashed] (4.45,-4)-- node [thick,midway,left,font=\small\sffamily,xshift=-.50em,yshift=.5em] {} (0.6,0);
\draw[->,thick] (6.45,-3)  -- node [thick,midway,right,font=\small\sffamily,xshift=-1.1em,yshift=.75em] {penalty kicker} (5.3,-3.5);
\draw[fill=red!20](5.65,.5) circle (0.20) node [black,yshift=-1.5cm] {};
\draw[rotate=-90,fill=red!20] (-.5, 5) ellipse (0.24cm and .55cm);
\begin{scope}[scale=.5,transform canvas={xshift=2.2cm,yshift=.45cm}]
\node [rotate=60,fill=red!20,transform shape,draw,rectangle,minimum width=.05cm,minimum height=1.5cm] {};
\end{scope}
\begin{scope}[scale=.5,transform canvas={xshift=2.2cm,yshift=.150cm}]
\node [rotate=90,fill=red!20,transform shape,draw,rectangle,minimum width=.05cm,minimum height=1.3cm] {};
\end{scope}
\begin{scope}[scale=.5,transform canvas={xshift=2.8cm,yshift=.4cm}]
\node [rotate=-70,fill=red!20,transform shape,draw,rectangle,minimum width=.05cm,minimum height=1cm] {};
\end{scope}
\begin{scope}[scale=.5,transform canvas={xshift=2.8cm,yshift=.2cm}]
\node [rotate=-90,fill=red!20,transform shape,draw,rectangle,minimum width=.05cm,minimum height=1cm] {};
\end{scope}
\draw[fill=red!20](4.8,-2.5) circle (0.250) node [black,yshift=-1.5cm] {};
\begin{scope}[scale=.5,transform canvas={xshift=2.2cm,yshift=-1.8cm}]
\node [rotate=-70,fill=red!20,transform shape,draw,rectangle,minimum width=.05cm,minimum height=1.6cm] {};
\end{scope}
\begin{scope}[scale=.5,transform canvas={xshift=2.5cm,yshift=-1.9cm}]
\node [rotate=10,fill=red!20,transform shape,draw,rectangle,minimum width=.05cm,minimum height=1.6cm] {};
\end{scope}
\begin{scope}[scale=.5,transform canvas={xshift=2.25cm,yshift=-1.5cm}]
\node [rotate=-30,fill=red!20,transform shape,draw,rectangle,minimum width=.05cm,minimum height=.95cm] {};
\end{scope}
\begin{scope}[scale=.5,transform canvas={xshift=2.5cm,yshift=-1.5cm}]
\node [rotate=-10,fill=red!20,transform shape,draw,rectangle,minimum width=.05cm,minimum height=.9cm] {};
\end{scope}
\draw[rotate=0,fill=red!20] (4.8, -3.2) ellipse (0.25cm and .55cm);
\node[above,font=\large\bfseries] at (3,3) {(a)};
\end{tikzpicture}
\begin{tikzpicture}[scale=.5]
\draw[line width=1mm,fill=LimeGreen] (0,-4.5) -- (12,-4.25) -- (8,.5) -- (-3,.50)--(0,-4.5) node{};
\draw[line width=.65mm][solid] (0,0)--(7.32,0)--(7.32,2.44)--(0,2.44)--(0,0);
\draw[very thick][fill=gray][dashed] (7.32,2.44)--(6.32,1.8);
\draw[very thick][dashed] (6.32,1.8)--(5.32,0);
\draw[very thick][dashed] (6,1.8)--(-1,1.8);
\draw[very thick][dashed] (0,2.44)--(-1,1.8);
\draw[very thick][dashed] (-1,1.8)--(-2,0);
\draw[very thick][dashed] (-2,0)--(0,0);
\draw[-,thick] (-1.95,.10)--(0,.10);
\draw[-,thick] (-1.85,.25)--(0,.250);
\draw[-,thick] (-1.78,.4)--(0,.4);
\draw[-,thick] (-1.65,.55)--(0,.55);
\draw[-,thick] (-1.62,.7)--(0,.7);
\draw[-,thick] (-1.52,.85)--(0,.85);
\draw[-,thick] (-1.46,1)--(0,1);
\draw[-,thick] (-1.35,1.15)--(0,1.15);
\draw[-,thick] (-1.28,1.3)--(0,1.3);
\draw[-,thick] (-1.18,1.48)--(0,1.48);
\draw[-,thick] (-1.1,1.65)--(0,1.65);
\draw[-,thick] (-.7,1.95)--(0,1.95);
\draw[-,thick] (-.55,2.1)--(0,2.1);
\draw[-,thick] (-.4,2.2)--(0,2.2);
\draw[-,thick] (-.3,2.3)--(0,2.3);
\draw[-,thick] (5.32,.10)--(7.32,.10);
\draw[-,thick] (5.42,.25)--(7.32,.250);
\draw[-,thick] (5.52,.4)--(7.32,.4);
\draw[-,thick] (5.62,.55)--(7.32,.55);%
\draw[-,thick] (5.72,.7)--(7.32,.7);
\draw[-,thick] (5.82,.85)--(7.32,.85);
\draw[-,thick] (5.92,1)--(7.32,1);
\draw[-,thick] (6.01,1.15)--(7.32,1.15);
\draw[-,thick] (6.10,1.3)--(7.32,1.3);
\draw[-,thick] (6.15,1.46)--(7.32,1.46);
\draw[-,thick] (6.28,1.65)--(7.32,1.65);%
\draw[-,thick] (6.41,1.8)--(7.32,1.8);
\draw[-,thick] (6.53,1.95)--(7.32,1.95);
\draw[-,thick] (6.85,2.08)--(7.32,2.08);
\draw[-,thick] (6.92,2.2)--(7.32,2.2);
\draw[-,thick] (7.15,2.3)--(7.32,2.3);
\draw[-,thick] (7.15,2.34)--(7.15,0);
\draw[-,thick] (7.0,2.2)--(7.0,0);
\draw[-,thick] (6.85,2.13)--(6.85,0);
\draw[-,thick] (6.7,2)--(6.7,0);
\draw[-,thick] (6.55,1.94)--(6.55,0);
\draw[-,thick] (6.4,1.83)--(6.4,0);
\draw[-,thick] (6.25,1.65)--(6.25,0);
\draw[-,thick] (6.1,1.4)--(6.1,0);
\draw[-,thick] (5.95,1.2)--(5.95,0);
\draw[-,thick] (5.8,.8)--(5.8,0);
\draw[-,thick] (5.65,.6)--(5.65,0);
\draw[-,thick] (5.45,.3)--(5.45,0);
\draw[-,thick] (-.15,2.4)--(-.15,0);
\draw[-,thick] (-.3,2.3)--(-.3,0);
\draw[-,thick] (-.45,2.13)--(-.45,0);
\draw[-,thick] (-.6,2.04)--(-.6,0);
\draw[-,thick] (-.75,1.94)--(-.75,0);
\draw[-,thick] (-.9,1.83)--(-.9,0); %
\draw[-,thick] (-1.05,1.65)--(-1.05,0);
\draw[-,thick] (-1.2,1.4)--(-1.2,0);
\draw[-,thick] (-1.35,1.2)--(-1.35,0);
\draw[-,thick] (-1.5,.85)--(-1.5,0);
\draw[-,thick] (-1.65,.56)--(-1.65,0);
\draw[-,thick] (-1.8,.39)--(-1.8,0);
\begin{scope}[scale=1.2,shift={(-.8,1.4)}]
      \fill[gray!30!white] (1.2,0.33) circle (0.32);
      \clip (1.2,0.33) circle (0.32);
      \fill[black] (1.06,0.30) -- (1.01,0.17) -- (1.14,0.08) -- (1.26,0.14) -- (1.20,0.28) -- cycle (1.37,0.14) -- (1.46,0.27) -- (1.59,0.27) -- (1.41,0.04) -- cycle (1.28,0.38) -- (1.22,0.52) -- (1.33,0.61) -- (1.45,0.51) -- (1.43,0.37) -- cycle (0.87,0.44) -- (1.02,0.40) -- (1.10,0.53) -- (1.07,0.62) -- (0.94,0.57) -- cycle;
\end{scope}
\draw[very thick](5.45,-4)[fill=black] circle (5pt) node{};
\draw[->,thick,dashed] (5.45,-4)-- node [thick,midway,left,font=\small\sffamily,xshift=.50em,yshift=3.5em] {} (.550,1.7);
\begin{scope}[scale=.5,transform canvas={xshift=.98cm,yshift=.35cm}]
\node [rotate=45,fill=red!20,transform shape,draw,rectangle,minimum width=.05cm,minimum height=2.3cm] {};
\end{scope}
\begin{scope}[scale=.5,transform canvas={xshift=.98cm,yshift=.250cm}]
\node [rotate=30,fill=red!20,transform shape,draw,rectangle,minimum width=.05cm,minimum height=2.3cm] {};
\end{scope}
\begin{scope}[scale=.5,transform canvas={xshift=.55cm,yshift=.85cm}]
\node [rotate=40,fill=red!20,transform shape,draw,rectangle,minimum width=.05cm,minimum height=1.85cm] {};
\end{scope}
\draw[fill=red!20](1,1.75) circle (0.250) node [black,yshift=-.5cm] {};
\begin{scope}[scale=.5,transform canvas={xshift=.5cm,yshift=.85cm}]
\node [rotate=50,fill=red!20,transform shape,draw,rectangle,minimum width=.05cm,minimum height=1.85cm] {};
\end{scope}
\draw[rotate=45,fill=red!20] (1.95, -.2) ellipse (0.2cm and .5cm);
\draw[fill=red!20](5.8,-2.5) circle (0.250) node [black,yshift=-1.5cm] {};
\begin{scope}[scale=.5,transform canvas={xshift=2.6cm,yshift=-1.8cm}]
\node [rotate=-70,fill=red!20,transform shape,draw,rectangle,minimum width=.05cm,minimum height=1.2cm] {};
\end{scope}
\begin{scope}[scale=.5,transform canvas={xshift=3cm,yshift=-1.9cm}]
\node [rotate=10,fill=red!20,transform shape,draw,rectangle,minimum width=.05cm,minimum height=1.8cm] {};
\end{scope}
\begin{scope}[scale=.5,transform canvas={xshift=2.75cm,yshift=-1.5cm}]
\node [rotate=-30,fill=red!20,transform shape,draw,rectangle,minimum width=.05cm,minimum height=.95cm] {};
\end{scope}
\begin{scope}[scale=.5,transform canvas={xshift=3cm,yshift=-1.5cm}]
\node [rotate=-10,fill=red!20,transform shape,draw,rectangle,minimum width=.05cm,minimum height=.9cm] {};
\end{scope}
\draw[rotate=0,fill=red!20] (5.8, -3.2) ellipse (0.25cm and 0.55cm);
\node[above,font=\large\bfseries] at (3,3) {(b)};
\end{tikzpicture}
\caption{(a): The goal scorer assigns ball (observation) to first cluster, but the goalkeeper assigns it to the third cluster. (b): Both, the goal scorer and goalkeeper assign the ball (observation) to seventh cluster.}
\label{fig3}
\end{figure}
\par
Based on the concept of clustering, we propose another tool called direction detection index (DDI) that is in fact a modification of the Rand index. We have
\begin{align}\label{DDI}
{\text{DDI}}=1-\sum_{i=1}^{r}\frac{1}{\underset{k=1,\cdots,c}{\max}~{d(\mu_{1i},\mu_{2k})}}\sum_{j=1}^{c}n_{ij} \times d
(\mu_{1i},\mu_{2j}),
\end{align}
where $n_{ij}$ (for $i=1,\cdots,r$ and $j=1,\cdots,c$) is represented in Table \ref{tab1} and $d(\mu_{1i},\mu_{2j})$ represents distance between centers of $i$-th cluster in model $M_1$ and $j$-th cluster in model $M_2$. The x-coordinate and y-coordinate of $\mu_{1i}$ (or $\mu_{2j}$) are given earlier in Subsection \ref{method}. For example, we have $\mu_{17}=(0,2.44)$ and $\mu_{26}=(7.32,1.22)$. Similar to the RI, we have $0\leq {\text{DDI}}\leq1$. For practical purposes, the limitations of RI measure can be circumvented by constructing a new measure that is minimum of RI and DDI. The new measure, called MRDI, is given by
\begin{align}\label{MRDI}
{\text{MRDI}}=\min\bigl\{{\text{RI}} , {\text{DDI}}\bigr\}.
\end{align}
\par We mention that the SV measure defined in (\ref{sv}) cannot evaluate the performance of the goalkeeper efficiently. As major weakness of the SV is its limitation in discriminating between two goalkeepers of unsuccessful savings. Although both goalkeepers are unsuccessful in saving the kicked ball, but clearly one who has detected the true cluster should be preferable to that one cannot. In what follows, we propose the goalkeeper's saving index (GSI) as a complementary point statistic for evaluating the goalkeeper's performance.
\begin{align}\label{GSI}
{\text{GSI}}=\min\biggl\{1, \max\Bigl[0,\frac{n_{s}+\omega_{e}\times (n_{ie}+n_{oe})-\omega_{d}\times (n_{id}+n_{od})}{n}\Bigr]\biggr\},
\end{align}
where
\begin{enumerate}[i]
\item $n=$ total number of penalty kicks,
\item $n_{s}=$ number of saved kicked balls inside the goal,
\item $n_{ie}=$ number of kicked balls inside the goal for which the goalkeeper detected the true cluster,
\item $n_{oe}=$ number of kicked balls outside the goal for which the goalkeeper detected the true cluster,
\item $n_{id}=$ number of kicked balls inside the goal for which the goalkeeper does not detect the true cluster,
\item $n_{od}=$ number of kicked balls outside the goal for which the goalkeeper does not detect the true cluster.
\item $0<\omega_{e}<0.5$: a quantity for rewarding the goalkeeper who detects the true cluster.
\item $0<\omega_{d}<0.5$: a quantity for penalizing the goalkeeper who dose not detect the true cluster.
\end{enumerate}
It is worthwhile to note that $n=n_{ie}+n_{id}+n_{oe}+n_{od}$ and hence $0\leq {\text{GSI}}\leq 1$. A goalkeeper with higher performance yields a larger ${\text{GSI}}$. Determining the quantities $\omega_{e}$ and $\omega_{e}$ is a technical task and must be determined by an expert. In general, it would be wise to say that $0<\omega_{d}<\omega_{e}<0.5$, if advantage of detecting the true cluster is more than the disadvantage of not detecting the true cluster. In this work, we suggest to set $\omega_{e}=0.3$ and $\omega_{e}=0.2$.
\section{Results}
This section has two parts. First, we investigate the performance of proposed measures RI given in (\ref{RI}), DDI given in (\ref{DDI}), and MRDI given in (\ref{MRDI}) through two artificial examples. The performance analysis of all proposed measures will be demonstrated in the second subsection by analyzing the real data correspond to four important international matches.
\subsection{Artificial data analysis}
Herein, we give two artificial examples for demonstrating the efficiency of the proposed tools for computing the performance of goalkeeper in detecting the direction of kicked ball.

\begin{enumerate}[\indent {}]
\item {\bf{Example 1}}: 
In the first example, we assume that there are ten kicked balls with true cluster model $M_{true}=\{1,1,3,3,1,2,1,2,8,9\}$. Moreover, there are two goalkeepers that propose cluster models represented as $M_{1}=\{3,1,3,3,1,2,1,2,3,4\}$ and $M_{2}=\{2,1,3,3,1,2,1,2,7,8\}$ for detecting the direction of ten kicks. The corresponding contingency tables are given in Appendix \ref{app.C}. For this example, the computed measures RI, DDI, and MRDI, indicate that the second goalkeeper shows superior performance than the first one as expected (see Table \ref{tab2}).
\item {\bf{Example 2}}: 
Herein, we assume that there are eight kicked balls with true cluster model $M_{true}=\{1,3,4,2,1,3,4,1\}$ and two goalkeepers proposing cluster models $M_{1}=\{1,1,3,3,1,3,4,2\}$ and $M_{2}=\{1,1,3,3,1,2,1,1\}$. 
In this example, the first goalkeeper outperforms the second one in terms of all three measures (see Table \ref{tab2}).
\end{enumerate}
\begin{table}[h]
\caption{Computed RI, DDI, and MRDI measures for the Example 1 and Example 2.}
\centering
{\begin{tabular}{ccccc}
\hline
\hline
\multirow{1}{*}{}&\multicolumn{1}{c}{cluster}&\multicolumn{3}{c}{measure}\\
\cline{3-5}
     & model&RI&DDI&MRDI\\
\hline
 \multirow{2}{*}{first Example}  &$M_{1}$ & 0.822& 0.708& 0.708 \\                                                                                             
\cline{3-5}                                                                                                                                                                          
                          				  &$M_{2}$ & 0.888& 0.821& 0.821 \\                                                                                              
                          				  \cline{1-5}                                                                                                                                                                          
 \multirow{2}{*}{second Example} &$M_{1}$&   0.678 &0.593  &0.593 \\                                                                                             
\cline{3-5}                                                                                                                                                                          
                          				  &$M_{2}$  &0.642  &0.572 & 0.572\\                                                                                              
\hline
\hline
\end{tabular}}
\label{tab2}
\end{table}
\subsection{Real data analysis}
Herein, we apply the measures RI, DDI, MRDI, and GSI, respectively given in (\ref{RI}), (\ref{DDI}), (\ref{MRDI}), and (\ref{GSI}), to compare the performance of goalkeepers during the penalty kicks recorded in four matches including: World cup 2006 between Argentina and Germany, UEFA Euro 2020 between Italy and England, COPA America 2024 between Canada and Venezuela, and World cup 2022 between Argentina and France. The results are summarized in Tables \ref{tab3}-\ref{tab6} in which the red-colored entries indicate that the kicked ball is outside of goal.
We record the following facts form analysis of Tables \ref{tab3}-\ref{tab6}.
\begin{enumerate}[i]
\item The DDI works well in all cases, but the RI sometimes fail to work well such as Table \ref{tab5} (Venezuela goalkeeper) and Table \ref{tab6} (France goalkeeper). This inefficiency of RI is due to small number of kicks.
\item For such cases as Tables \ref{tab4}-\ref{tab6} in which we have red-colored records, or the cases in which goalkeeper detects true cluster with allowed goals (such as Table \ref{tab3} - Germany goalkeeper case), then the GSI can be recognized as a fair measure to evaluate the goalkeeper's performance.
\end{enumerate}
\section{Discussion}
Although the RI index has a strong theoretical background, but there follows a listing of its limitations in evaluating the performance of goalkeeper.
\begin{enumerate}[i]
\item Suppose we have four kicked balls at clusters 1, 7, 1, and 7 for which the true cluster becomes $M_{true}=\{1,7,1,7\}$. Furthermore, let the first an second goalkeepers suggest models $M_{1}=\{1,4,1,7\}$ and $M_{2}=\{1,3,1,7\}$, respectively. Obviously, both goalkeepers work differently just in detecting the direction of the second kicked ball. It sounds that the first goalkeeper works better than the second one since she/he assigns the second kicked ball to the fourth cluster that is closer to the true cluster 7 than the third cluster detected by the second goalkeeper. Unfortunately, the RI becomes 0.833 for both goalkeeper. This is while the DDI for first and second goalkeepers are 0.960 and 0.750, respectively.
\item The RI is not sensitive to the cluster label switching. For example, if we have two cluster models as $M_{true}=\{1,1,3,3\}$ and $M_{1}=\{3,3,1,1\}$. As it is seen, the goalkeeper with model $M_{1}$ detects all clusters incorrectly, but not surprisingly we have ${\text{RI}}=1$. The DDI for this case is 0.051.
\item The RI may fail to work well when number of kicked balls (observations) is small. This weakness is a major obstacle for using the RI since usually there are limited number of penalty kicks if the match result needs to be determined in the knock-out stage.
\end{enumerate}
All settings suggested in this work for constructing RI, DDI, MRDI (including number, centers, and shapes of clusters), and GSI (including $\omega_{e}$ and $\omega_{d}$) measures may be changed appropriately in practice. We record the following facts for completeness.
\begin{enumerate}[i]
\item We supposed that the goal area is partitioned into nine zones, but the number of zones can be changed if desired. 
\item The proposed measures in this work have been designated for penalty kicks, but may be applicable for the free kicks and furthermore for other games such as handball and indoor soccer.
\item If the goalkeeper has no movement when receiving the kicked ball, then we assumed that the detected cluster by goalkeeper is 5.
\item The shape of clusters may be changed appropriately. For example, as shown by Figure \ref{fig2} (b), the clusters can have elliptical forms with different sizes.
\item The nearest hand of goalkeeper to the kicked ball determines the goalkeeper's detected cluster. For example, if the fourth zone is true cluster and the right (left) hand of goalkeeper is closer to the kicked ball, but his left (right) hand lies at the first zone, then the fourth zone is considered as the cluster detected by goalkeeper. 
\end{enumerate}
The pertinent \verb+R+ code for computing the RI and DDI are given in Appendices \ref{app.A} and \ref{app.B}, respectively. 
\section{Conclusion}
The performance of goalkeepers in detecting the direction of penalty kicks to the goal have been evaluated an then compared by four measures. The first three measures including the Rand index (RI), direction detection index (DDI), minimum of RI, and DDI (MRDI) have been constructed based on the concept of clustering. For computing these three measures, it has been assumed that the goal area is partitioned into nine zones, but the number of zones can be changed if desired. The proposed DDI measure works well for evaluating and comparing the performance of goalkeepers. The fourth proposed measure, that is the goalkeeper saving index (GSI), is in fact a modification of the saving percentage (SV) statistic. The GSI accounts the reward (or penalty) for the goalkeeper's decision in detecting the direction of the kicked ball correctly (or incorrectly). Our findings suggest that the all four measures are useful statistics for evaluating and comparing the performance of goalkeepers. All measures have been proposed in this work are applicable for free kicks or other games such as handball and indoor soccer.
\begin{itemize}
\item {\bf {Funding details:}}~~There is no fund for this work.
\item {\bf {Conflicts of Interest:}}~~The author declares that he has no conflicts of interest.
\item {\bf {Total number of works in abstract, tables and figures:}}~~The manuscript includes 149 words for Abstract, 6 tables and 3 figures.
\item {\bf {Data availability statement:}}~~The data that support the findings of this study are openly available in World wide web.
\end{itemize}

\newpage{}
\begin{table}[!h]
\caption{World cup 2006 match: Argentina (goalkeeper {\bf{Leo Franco}}) versus Germany (goalkeeper {\bf{Jens Lehmann}}).}
\centering
{\begin{tabular}{lllll}
\hline
\hline
                                           shoot    &Argentina           & true               &Germany goalkeeper &goal         \\
                                           number &penalty kicker    & cluster       &clustering                 &result     \\
\hline
                                            1 &Julio Cruz                   &7                     &7                             &allowed \\                                                                                             
                                           2 &Roberto Ayala             &3                     &3                             &saved     \\                                                                                             
                                           3 &Maximiliano Rodriguez & 1                    &1                             &allowed  \\                                                                                             
                                           4 &Esteban Cambiasso     & 3                    &3                             &saved     \\                                                                                             
\hline     
                                           shoot    &Germany& true       & Argentina goalkeeper                   &goal       \\
                                           number &penalty kicker & cluster&clustering                                 &result     \\
\hline
                                            1 &Oliver Neuville           & 6                    &6                              &allowed  \\                                                                                             
                                            2 &Michael Ballack          & 8                   &5                              &allowed  \\                                                                                             
                                            3 &Lukas Podolski           & 3                   &2                              &allowed  \\                                                                                             
                                            4 &Tim Borowski             & 3                   &2                              &allowed  \\
                                           \hline
                                           goalkeeper stats   &         &                      &                                &\\                                                                                                                                                                                     
                                           \multicolumn{1}{r}{Lehmann:}& \multicolumn{4}{l}{RI=1, DDI=1.000, MRDI=1.000, SV=0.500, GSI=0.800}\\                                                                                                                                                                                     
                                           \multicolumn{1}{r}{Franco:}     & \multicolumn{4}{l}{RI=1, DDI=0.693, MRDI=0.693, SV=0.000, GSI=0.250}\\                                                                                             
\hline
\hline
\end{tabular}}
\label{tab3}
\end{table}
\begin{table}[!h]
\caption{UEFA Euro 2020 final match: Italy (goalkeeper {\bf{Gianluigi Donnarumma}}) versus England (goalkeeper {\bf{Jordan Pickford}}).}
\centering
{\begin{tabular}{lllll}
\hline
\hline
                 shoot    &England                     & true                &Italy goalkeeper        &goal     \\
                 number &penalty kicker             & cluster         &clustering                 &result   \\
\hline
                           1 &Harry Kane                 &1                     &1                             &allowed \\                                                                                             
                           2 &Harry Maguire             &9                     &1                             &allowed \\                                                                                             
                           3 &Marcus Rashford         &{\bf{\color{red}{1}}}&3                     &saved  \\                                                                                             
                           4 &Jadon Sancho             & 3                    &3                             &saved   \\   
                           5 &Bukayo Saka               & 3                    &3                             &saved   \\   
\hline     
                  shoot    &Italy                           & true               & England goalkeeper  &goal       \\
                  number &goal soccer                 & cluster        &clustering                 &result     \\
\hline
                           1 &Domenico Berardi      & 1                   &3                               &allowed  \\                                                                                             
                           2 &Andrea Belotti           & 3                   &3                               &saved  \\                                                                                             
                           3 &Leonardo Bonucci      & 2                   &2                               &allowed  \\                                                                                             
                           4 &Federico Bernardeschi& 2                   &1                               &allowed  \\
                           5 &Jorginho                    & 1                  &1                               &saved  \\
\hline
                           goalkeeper stats   &         &                      &                                &\\                                                                                                                                                                                     
                           \multicolumn{1}{r}{Pickford:}& \multicolumn{4}{l}{RI=0.600, DDI=0.643, MRDI=0.600, SV=0.400, GSI=0.540}\\                                                                                                                                                                                     
                          \multicolumn{1}{r}{Donnarumma:}     & \multicolumn{4}{l}{RI=0.600, DDI=0.610, MRDI=0.600, SV=0.600, GSI=0.540}\\                                                                                             
\hline
\hline
\end{tabular}}
\label{tab4}
\end{table}
\begin{table}[!h]
\caption{COPA America 2024 match: Canada (goalkeeper {\bf{Maxime Crepeau}}) versus Venezuela (goalkeeper {\bf{Rafael Romero}}).}
\centering
{\begin{tabular}{lllll}
\hline
\hline
                                           shoot    &Canada                & true               &Venezuela goalkeeper &goal         \\
                                           number &penalty kicker           & cluster       &clustering                 &result     \\                                        
\hline
                                            1 &Jonathan David          & 6                   &6                              &allowed  \\                                                                                             
                                            2 &Liam Millar                 & {\bf{\color{red}{9}}}&1                   &saved  \\                                                                                             
                                            3 &Moïse Bombito           & 3                   &1                              &allowed  \\                                                                                             
                                            4 &Stephen Eustáquio     &2                    &2                              &saved  \\
                                            5 &Alphonso Davies        &9                     &1                &allowed  \\
                                            6 &Ismaël Koné              &1                     &3                              &allowed   \\   
\hline     
                                           shoot    &Venezuelan& true       & Canada goalkeeper                    &goal       \\
                                           number &penalty kicker & cluster&clustering                                 &result     \\
\hline                     
                                            1 &Salomón Rondón          &3                     &1                           &allowed \\                                                                                             
                                            2 &Yangel Herrera             &{\bf{\color{red}{1}}}&3                  &saved \\                                                                                             
                                            3 &Tomás Rincón              & 2                    &1                           &allowed  \\                                                                                             
                                            4 &Jefferson Savarino        & 4                    &4                           &saved   \\   
                                            5 &Jhonder Cádiz              & 3                    &1                           &allowed   \\   
                                            6 &Wilker Ángel                & 3                    &3                           &saved   \\   
                                           \hline
                                           goalkeeper stats   &         &                      &                                &\\                                                                                                                                                                                     
                                           \multicolumn{1}{r}{Crepeau:}& \multicolumn{4}{l}{RI=0.666, DDI=0.386, MRDI=0.386, SV=0.500, GSI=0.466}\\                                                                                                                                                                                     
                                           \multicolumn{1}{r}{Romero:}     & \multicolumn{4}{l}{RI=0.866, DDI=0.350, MRDI=0.350, SV=0.333, GSI=0.350}\\                                                                                             
\hline
\hline
\end{tabular}}
\label{tab5}
\end{table}
\begin{table}[!h]
\caption{World cup 2022 match: Argentina (goalkeeper {\bf{Emiliano Martinez}}) versus France (goalkeeper {\bf{Hugo Lloris}}).}
\centering
{\begin{tabular}{lllll}
\hline
\hline
                                           shoot    &Argentina        & true               &France goalkeeper &goal       \\
                                           number &penalty kicker  & cluster           &clustering                 &result     \\
\hline
                                            1 &Lionel Messi            &1                    &1                             &allowed \\                                                                                             
                                            2 &Paulo Dybala          &2                    &3                              &allowed  \\                                                                                             
                                            3 &Leandro Paredes     &1                    &1                              &allowed  \\                                                                                             
                                            4 &Gonzalo Montiel      &1                    &3                              &allowed     \\                                                                                             
\hline     
                                           shoot    &Germany& true       & Argentina goalkeeper               &goal       \\
                                           number &penalty kicker & cluster&clustering                             &result     \\
\hline
                                            1 &Kylian Mbapp{\'e}   &2                    &2                              &allowed  \\                                                                                             
                                            2 &Kingsley Coman      &1                    &1                              &saved  \\                                                                                             
                                            3 &Aur{\'e}lien Tchouam{\'e}ni&{\bf{\color{red}{1}}}&1 &saved  \\                                                                                             
                                            4 &Randal Muani         &2                    &1                               &allowed  \\
                                           \hline
                                           goalkeeper stats   &           &                      &                                &\\                                                                                                                                                                                     
                                           \multicolumn{1}{r}{Martinez:}& \multicolumn{4}{l}{RI=1, DDI=1.000, MRDI=1.000, SV=0.500, GSI=0.450}\\                                                                                                                                                                                     
                                           \multicolumn{1}{r}{Lloris:}     & \multicolumn{4}{l}{RI=1, DDI=0.693, MRDI=0.693, SV=0.000, GSI=0.250}\\                                                                                             
\hline
\hline
\end{tabular}}
\label{tab6}
\end{table}
\clearpage 
\bibliographystyle{spphys}       
\bibliography{ref1}
\appendix
\section{R code for computing the Rand index}\label{app.A}
\begin{verbatim}
############## R function for computing the RI ###########
RI <- function(x, y){
n <- length(x)
dx <- as.matrix(dist(x, diag=T, upper=T), nrow=n, ncol=n)
dy <- as.matrix(dist(y, diag=T, upper=T), nrow=n, ncol=n)
ri <- 1 - sum( abs( sign(dx) - sign(dy) ) )/( n*(n - 1) )
return(ri)
}
##########################################################
\end{verbatim}
\section{R code for computing the direction detection index}\label{app.B}
\begin{verbatim}
############## R function for computing the DDI ##########
DDI <- function(x, y, meter = "euclidean")
{
	if( meter != "euclidean" & meter != "maximum" & meter !=
   "manhattan" & meter != "canberra" & meter != "binary" &
    meter != "minkowski")
	stop("method spelling is not correct. Method must be one 
	      of euclidean, maximum, manhattan, canberra, binary,
	      and minkowski.")
	L <- 7.320 # length of crossbar
	A <- 2.44  # length of goalpost 
	n.x <- 3   # number of segments in crossbar
	n.y <- 3   # number of segments in goalpost
	level.x <- as.numeric( levels( as.factor(x) ) )
	level.y <- as.numeric( levels( as.factor(y) ) )
	xg <- seq(0, L, length = n.x)
	yg <- seq(0, A, length = n.y)
	M <- cbind( rep(xg, n.y), rep(yg, each = length(xg) ) )
	dis <- as.matrix(dist(M, method=meter, diag=T, upper=T), 
	nrow = n.x*n.y, ncol = n.x*n.y )
	n.ij <- matrix(0, nrow = n.x*n.y, ncol = n.x*n.y)
	tab <- table(x, y)
	n.ij[level.x, level.y] <- tab
	ddi <- 1 - sum( n.ij*dis/apply(dis, 2, max) )/sum(n.ij)
return(ddi)
}
##########################################################
\end{verbatim}
\clearpage
\section{Contingency tables for the first example}\label{app.C}
{\small{\begin{verbatim}
       M.true                                           
  M1  [,1] [,2] [,3] [,4] [,5] [,6] [,7] [,8] [,9]        
 [1,]    3    0    1    0    0    0    0    0    0        
 [2,]    0    2    0    0    0    0    0    0    0        
 [3,]    0    0    2    0    0    0    0    0    0        
 [4,]    0    0    0    0    0    0    0    0    0        
 [5,]    0    0    0    0    0    0    0    0    0        
 [6,]    0    0    0    0    0    0    0    0    0        
 [7,]    0    0    0    0    0    0    0    0    0        
 [8,]    0    0    1    0    0    0    0    0    0        
 [9,]    0    0    0    1    0    0    0    0    0

       M.true 
  M2  [,1] [,2] [,3] [,4] [,5] [,6] [,7] [,8] [,9] 
 [1,]    3    1    0    0    0    0    0    0    0 
 [2,]    0    2    0    0    0    0    0    0    0 
 [3,]    0    0    2    0    0    0    0    0    0 
 [4,]    0    0    0    0    0    0    0    0    0 
 [5,]    0    0    0    0    0    0    0    0    0 
 [6,]    0    0    0    0    0    0    0    0    0 
 [7,]    0    0    0    0    0    0    0    0    0 
 [8,]    0    0    0    0    0    0    1    0    0 
 [9,]    0    0    0    0    0    0    0    1    0 
  \end{verbatim}
  }}
\end{document}